\documentstyle[preprint,aps]{revtex}

\def\p{{\mbox{\boldmath$p$}}}
\def\r{{\mbox{\boldmath$r$}}}
\def\q{{\mbox{\boldmath$q$}}}
\def\s{{\mbox{\boldmath$s$}}}
\def\x{{\mbox{\boldmath$x$}}}

\def\dern0{\frac{\partial n^0}{\partial\varepsilon_p}}

\newcommand{\beln}[1]{\\@#1 \begin{equation}\label{#1}}
\newcommand{\bel}[1]{\begin{equation}\label{#1}}
\newcommand{\ee}{\end{equation}}

\newcommand{\simless}{\buildrel < \over \sim}

\begin{document}
%%%%%%%%%%%%%%%%%%%%%%%%%%%%%%%%%%%%%%%%%%%%%%%%%%%%%%%%%%%%%%%
\draft

\title{
\begin{flushright}
{\rm MSUCL-1047}
\end{flushright}
Fragments in Gaussian Wave-Packet Dynamics\\ with
and without Correlations}

\author{Dieter Kiderlen\thanks{e-mail:
kiderlen@nscl.msu.edu}
 and Pawe{\l} Danielewicz\thanks{e-mail:
danielewicz@nscl.msu.edu}
\\
National Superconducting Cyclotron Laboratory and\\
Department of Physics and Astronomy, Michigan State University,
\\
East Lansing, Michigan 48824, USA\\
}

\date{\today}
\maketitle

\begin{abstract}
Generalization of Gaussian trial wave functions in quantum
molecular dynamics models is introduced, which
allows for
long-range correlations characteristic for composite nuclear
fragments. We~demonstrate a~significant improvement in the
description of light fragments with correlations.  Utilizing
either type of Gaussian wave functions, with or without
correlations, however, we~find that we cannot describe
fragment formation in a~dynamic situation.  Composite fragments
are only produced in simulations if they are present as
clusters in the substructure of original nuclei.
The~difficulty is traced to the delocalization of wave
functions during emission.  Composite fragments are produced
abundantly in the Gaussian molecular dynamics in the limit
$\hbar \rightarrow 0$.
\end{abstract}

\pacs{PACS numbers: 24.10.-i, 25.70.-z, 25.70.Pq\\[-1.5ex]
\noindent
keywords: wave-packet dynamics, fragment production,
correlations, FMD}

\section{INTRODUCTION}

The theoretical description of heavy-ion collisions at intermediate
beam energies, 10 MeV $\simless E/A \simless $ 150 MeV, is still in
an unsatisfactory state.  Different factors contribute to that
situation.  Thus, for one, the nucleon excitation energies are low
and in that comparable to nucleon localization energies, indicating
a~likely importance of the quantal effects.  With the change in the
beam energy in the discussed range, the dynamics changes from that
dominated by the mean field to that dominated by collisions (as
evidenced in the appearance of the flow balance-energy).
As~excitation energies grow, they begin to exceed the average
binding energies per nucleon and, within the mentioned range of
$E/A$, a~massive production of intermediate-mass fragments (IMF)
takes place \cite{hir94}.  The production, in terms of IMF
multiplicity or total mass that IMF carry, maximizes at
$E/A \sim 75$ MeV.  The~description of the intermediate and light
fragment production is beyond the capability of common
single-particle models of collisions \cite{ber88}.  The
single-particle models with fluctuating forces  \cite{ayi90,ran90}
can describe fragment production, but miss the shell effects and
the discreteness of the mass and charge numbers.
The~involved limitation is recognized once one realizes that,
in the very central Au + Au collisions at 100~MeV/nucleon,
the probability for a proton to come out from the reaction as
a~constituent of an~$\alpha$ particle is close \cite{tsa94}
to~50\%.  Within the Boltzmann-Langevin model with the
fluctuating forces \cite{ayi90,ran90}, the~$\alpha$ particle
plays no distinguished role.
Statistical models \cite{bon85,gro86} account for the shell
effects, but miss the reaction dynamics.
The~importance of the dynamics is seen, in~particular, in
the fact that the~collective outward flow energy in the
reactions is comparable to the thermal energy.

Given the above situation, the quantum molecular
model
proposed in recent years for the reactions \cite{fel90} was
met with quite some expectations.
In this model, the~quantal wave function for a~reacting system
is represented as a~product of Gaussian wave-packets for
individual nucleons.
The packets have dynamic centers, phases, and widths.
The~parameters obey equations of motion following from
a~variational principle.
The~model \cite{fel90} accounts for shell effects and has been
shown to describe the evaporation of individual nucleons from
excited nuclei \cite{fel95}.
As~nucleons leave a~nucleus, their wave-packets become
completely delocalized.
The~packets are used in this model in favor of the Hartree-Fock
wave-functions, because with the packets one expects to
describe fluctuations.  Note that
initialization of the classical Vlasov equation with a set of
$\delta$-functions leads to the same results as the exact
equations of motion.
The~set of wave-packets is the closest approximation to the set
of $\delta$-functions that one can get quantally.
Other molecular models in use might be considered amended or
simplified versions of \cite{fel90}.
Both in the model \cite{fel90} (termed~FMD) and in \cite{ono92}
(termed~AMD), the overall wave function is  antisymmetrized.
However, in \cite{ono92} the dynamics of the wave-packet width
is suppressed.
In \cite{mar96} (EQMD) the width is dynamic, but the effects of
antisymmetrization are accounted for only approximately, using
a~Pauli pseudopotential.
In \cite{aic86} and \cite{mar90} (QMD) neither the widths are
dynamic nor the antisymmetrization is carried out explicitly.
These models are the most classical within the class.
In~the models AMD, EQMD, and QMD, collisions between wave
packets take place, on top of the wave-packet dynamics
obtained or attributable to the~variational principle for wave
functions.

Within the molecular models, fragment production has been studied,
in quite some detail, in the most classical of the class, the QMD
model \cite{boh91,gos95}.  Within the FMD model, the fragment
emission has been observed in the calculations of reactions
involving light nuclei~\cite{fel95}.
In~the FMD model a~specific problem arises concerning the
emission of fragments.
Inside a~fragment, the~constituent nucleons are localized in
the relative separations.  When the wave function is a product
of single-particle wave functions, this implies a localization
of the fragment center of mass.
While low-energy nucleons escaping from a~nucleus become
delocalized, getting rid of the kinetic energy associated with
their localization \cite{fel95}, this cannot be the case for
the fragments.
The~c.m.~localization energies e.g.~in the range of $A=($2--4)
nuclei, are within $\Delta E \sim (10-30)$~MeV and, given that
the
temperatures in excited nuclei in reactions could be as low as
(5--10)~MeV, they could result in significant thermal
penalty-factors for emission, $\rm{e}^{-\Delta E/T}$.

Generally, within FMD and~EQMD the
kinetic energy of the localization of the center of mass is not
conserved and
may be transformed into the internal energy, in
a~translationally
invariant situation.  While this need not be a~problem for
large fragments with a~small localization energy compared
to the fragment total energy, for light fragments it can mean
that their excitation energies cannot be determined.

In this paper, we investigate fragment production within
a~model of the~FMD type.
Given the problems associated with the localization, mentioned
above,
we consider the wave functions, of a~Gaussian form, that are
products of wave functions for individual nucleons and also
such that allow for the correlations between nucleons within
fragments, with a decoupling of the fragment center of mass
motion.
At present, effects of antisymmetrization are accounted for
approximately only using a~Pauli potential.

The outline of the paper is as follows.
Section \ref{trial} describes
the trial wave functions for the Schwinger variational
principle.
The~equations of motion
following from that principle are discussed in
Sec.~\ref{eom}.  Section~\ref{hamiltonian} discusses
a~choice of the hamiltonian which permits
an~analytic calculation of the expectation values in the
equations.  Our results
on fragment dynamics and production are reported in
Sec.~\ref{solution}
and the conclusions are given in Sec.~\ref{summary}.

\section{TRIAL WAVE FUNCTION}
\label{trial}

The trial wave-function for the Schwinger variational principle
is taken in the form
\bel{trialwf}
\langle \x_1,\dots,\x_N |\Psi \rangle ={\cal N}
\exp\Big(\phi( \x_1,\dots,\x_N)\Big)\, \chi \, ,
\ee
where $\chi$ is a normalized spin-isospin wave function.
The~argument in the unnormalized exponential wave function is
\bel{expon}
\phi( \x_1,\dots,\x_N )=-A_{ij}\,(\x_i
-\r_i) \cdot (\x_j-\r_j) +i \, \p_i \, \x_i.
\ee
Here,
$\x_i$ denotes the position vector
of particle~$i$.  The~repeated particle indices indicate
summation.  In~absence of antisymmetrization,
from (\ref{trialwf}) and (\ref{expon}) one finds that the
normalization constant
is equal to
${\cal N}= ({\rm det}(2 \, {\rm Re}\,A)/\pi^{N})^{3/4}$,
where $N$ is the particle number.

The parameters of the spatial wave function, which depend on
time, include the elements of the complex symmetric $3N \times
3N$ matrix $A$ with a positive definite real part, and further
the components of the $2N$ real vectors $\lbrace \r_i \rbrace
$ and $\lbrace \p_i \rbrace$.  Below, we
shall use $q_{\nu}$ to indicate any member of a~set
of the dynamic parameters.
A~convenient mapping of the set follows using
$q_{i(i-1)/2+j}={\rm Re}\,A_{ij}$ and
$q_{N(N+1)/2+i(i-1)/2+j}={\rm Im}\,A_{ij}$, where in the both
equations $j\le i$, and further using
$q_{N(N+1)+3(i-1)+a}=(\r_i)_a$
and $q_{N(N+4)+3(i-1)+a}=(\p_i)_a$.
The time-dependent parameters relate to different expectation
values with
\bel{av}
\langle \x_i \rangle = \r_i \, , \qquad \qquad
\langle -i \, \nabla_i \rangle = \p_i \, ,
\ee
\bel{flucr}
\langle (\x_i)_a \, (\x_j)_b \rangle =
(\r_i)_a \, (\r_j)_b +{ \delta_{ab} \over 4} \, ({\rm
Re}A)^{-1}_{ij}  \ee
and
\bel{flucp}
\langle i(\nabla_i)_a \, i(\nabla_j)_b \rangle =(\p_i)_a \,
(\p_j)_b+
\delta_{ab} \, (A^\ast \, ({\rm Re}\,A)^{-1} \, A)_{ij} \, ,
\ee
where $a$ and $b$ are indices for carthesian coordinates.
Given (\ref{av})--(\ref{flucp}), the parameters $\r_i$ and
$\p_i$ are referred to, further,
as the centroid and momentum of a~particle $i$,
respectively, and $A$ is referred to as the~width matrix.

The form (\ref{trialwf}) includes the special
case of the~width diagonal in particle indices,
$A_{ij}=\delta_{ij} \, A_{i}$.
The wave function in that case reduces to a~product of
single-nucleon wave functions such as utilized in
the FMD or EQMD calculations.  To~illustrate the advantage
that~(\ref{trialwf})  with~(\ref{expon})  offers, let us consider
a~deuteron.  The~parametrization allows for the wave function
of the form
\bel{deuwf}
\langle \x_1, \x_N |\Psi \rangle ={\cal N}
\exp\Big(-A_{cm} (\x_1 + \x_2)^2/4 - A_{rel} (\x_1 - \x_2)^2
\Big)\, \chi \, ,
\ee
where $A_{cm} = 2(A_{11}+A_{12})$, $A_{rel} =
(A_{11}-A_{12})/2$, and $A_{11}=A_{22}$.  The~magnitude of
$A_{rel}$ can be adjusted to reproduce the r.m.s.\ radius of
the deuteron and $A_{cm}$ in the above may
take on arbitrarily low values corresponding
to a~delocalized deuteron, as expected in the emission in
reactions.
FMD and EQMD parametrizations
more standard than ours,
with vanishing off-diagonal terms in~$A$, on~the other hand,
permit only strongly localized deuterons, as~$A_{cm}=4 A_{rel}$
for~$A_{12}=0$.

\section{EQUATIONS OF MOTION}
\label{eom}

The wave function (\ref{trialwf}) depends on time only
indirectly through the parameters $q_\mu$.
In this section, we shall obtain equations of motion (eom)
that govern the behavior of these parameters.

On writing the time-dependent variational principle in the form
$\delta\int_{t_1}^{t_2} {\cal L}(\dot q_\mu,q_\mu) dt=0$,
the eom for $q_\mu$ follow (formally) as Lagrange-Euler
equations,
${d\over dt}{\partial {\cal L}\over\partial \dot q_\mu}-
{\partial {\cal L}\over\partial q_\mu}=0$.
Using
${\cal L}= \langle\Psi\vert i\hbar{d\over dt} -H\vert\Psi \rangle$
for the Lagrange function, one obtains, see \cite{fel95},
\bel{eomimpl}
{\cal A}_{\nu\mu} \, \dot q_\mu = -{\partial\over\partial
q_\nu} \, \langle H \rangle \, ,
\ee
where the matrix ${\cal A}$, multiplying the time derivatives,
is skew symmetric.
This matrix is related to the overlap of the derivatives of the wave
function with respect to the parameters:
\bel{skewmat}
{\cal A}_{\nu\mu}= 2\, {\rm Im}\left\langle
{\partial\over\partial
q_\nu} \Psi \, \Bigg\vert \, {\partial\over\partial q_\mu}
\Psi\right\rangle \, .
\ee
Given the wave function of the form (\ref{trialwf}), one finds
for $\cal A$
\bel{skewalt}
{\cal A}_{\nu\mu}= 2 \, {\rm Im}\left(
\left[{\partial\phi^\ast\over\partial q_\nu} \,
\Big( {\partial\phi\over\partial q_\mu}-
\left[{\partial\phi\over\partial q_\mu}\right]
{\cal N}^2\Big)\right]{\cal N}^2
+\left\lbrace{\partial\chi\over\partial q_\nu} \, \right\vert
(1-\vert\chi\rbrace\lbrace\chi\vert) \, \left\vert
{\partial\chi\over\partial q_\mu}\right\rbrace \right) \, ,
\ee
where the square brackets stand for
$[O]= \int \Pi_{l=1}^N \, d\x_l \exp{(\phi^\ast)}
O \exp{(\phi)}$. From (\ref{skewalt}), one can see
that $\cal A$ does not couple parameters describing spin-isospin
degrees, with those in the spatial wave function.
For an~interaction diagonal in spin and isospin, considered
below, this
implies that the spin-isospin wave-function does not depend on
time.
Correspondingly, the~spin-isospin wave functions will be
largely disregarded in the further discussion.

The explicit expression for $\cal A$, obtained using
(\ref{expon}), is
\bel{skewexpl}
{\cal A}_{\nu\mu}=
{3\over 4} \, ({\rm Re}\,A)^{-1}_{in} \, ({\rm
Re}\,A)^{-1}_{jm}
\, {\partial \, {\rm Re}\,A_{ij}\over\partial q_\nu} \,
{\partial \, {\rm Im}\,A_{nm}\over\partial q_\mu} +
{\partial\r_i\over\partial q_\nu} \,
{\partial\p_i\over\partial q_\mu}- (\nu\leftrightarrow\mu) \, .
\ee
Since $A$ is independent of $\r_i$ and $\p_i$, the eom for
particle centroids and momenta take on the form of the
Hamilton's
equations with the expectation value of the Hamilton operator
in these equations playing the role of a~classical Hamiltonian.
As far as the width is concerned, for a practical solution of
the eom, it is necessary to invert the~$\cal A$-matrix.  When
represented in the space of the real and imaginary elements
of~$A$, the matrix~$\cal A$ acquires a~$2 \times 2$ block
structure with the blocks on the diagonal vanishing and
those
off-diagonal (coupling the real and imaginary parts
of~$A$) given by~$\pm {\cal B}$.  Fortunately,
the~quadratic matrix~$\cal B$, of a~dimension $N(N+1)/2$,
can be easily inverted analytically with the result
\bel{invcalA}
{\cal B}^{-1}_{{n(n-1)\over 2}+m,{i(i-1)\over 2}+j}=
{2\over 3} \, ( {\rm Re}\,A_{ni}\, {\rm Re}\,A_{mj}+
{\rm Re}\,A_{nj}\, {\rm Re}\,A_{mi}) \, ,
\ee
where $1\le m \le n\le N$, $1\le j \le i\le N$. The
equations of motion for the parameters -- centroids, momenta,
and matrix elements of~$A$ -- take then on the form
\bel{eomexpl2}
\dot \r_i = {\partial \langle H \rangle \over\partial\p_i} \,
,\qquad \, \dot \p_i = -{\partial \langle H
\rangle\over\partial\r_i} \, ,
\ee
\bel{eomexpl1}
{d\over dt} \, {\rm Re}\,A_{ij}= \sum_{m\le n}
{\cal B}^{-1}_{{i(i-1)\over 2}+j,{n(n-1)\over 2}+m} \,
{\partial \langle H \rangle \over\partial \, {\rm Im}\,A_{nm}}
\, ,\qquad \,
{d\over dt} \, {\rm Im}\,A=-{\cal B}^{-1} \, {\partial
\langle H \rangle \over\partial \, {\rm Re}\,A} \, .
\ee
The indices in the second equation in~(\ref{eomexpl1}) should
be handled in the same manner as in the first of the equations
where they are written out explicitly.  In~evaluating the
derivatives in~(\ref{eomexpl1}), the~elements ${\rm Re} \,
A_{nm}$ and ${\rm Re} \, A_{mn}$, and ${\rm Im} \, A_{nm}$
and ${\rm Im} \, A_{mn}$, respectively, should be treated as
identical.

Given~(\ref{invcalA}), the~contributions from the kinetic
energy to the~eom~(\ref{eomexpl1}) for the width matrix
may be simplified into
\bel{eomfree}
{\cal B}^{-1} \, {\partial\langle T\rangle\over\partial \, {\rm
Im}\, A}= {2\hbar\over m} \, {\rm Im}\, A^2, \qquad\qquad
{\cal B}^{-1} \, {\partial\langle T\rangle\over\partial \, {\rm
Re}\, A}= {2\hbar\over m} \, {\rm Re}\, A^2 \, .
\ee
For the free particles, eom may be then represented as
$\dot A= -2 i \hbar A^2/m$, with the solution
$A(t)=A_0 \, (1+2i\hbar(t-t_0)A_0/m)^{-1}$.  This means that,
for free particles, a~width matrix diagonal initially in
certain directions  in the~$N$-particle space  will stay
diagonal
in these directions.  Contributions to the kinetic energy
associated with different directions, proportional to
$|A(t)|^2/{\rm Re} \, A(t) = |A_0|^2/{\rm Re} \, A_0$, stay
constant as a~function of time.  For~interacting particles,
clearly, interaction terms appear on the r.h.s.~of the eom for
the width.  These terms can, generally, cause violations of the
conservation law for the c.m.~kinetic energy, when the width
matrix is constrained to the~diagonal form.

\section{HAMILTONIAN}
\label{hamiltonian}

The eom (\ref{eomexpl2}) and (\ref{eomexpl1}) express the time
derivatives of the parameters $q_\mu$ in the wave function as
the linear combinations of the derivatives of the expectation
value of the Hamiltonian, with respect to~$q_\mu$.
We~shall now calculate the Hamiltonian expectation value in
terms of the parameters.

The expectation value of the kinetic energy,
$\langle T\rangle=-{\hbar^2\over 2m}\sum_i \langle\nabla^2_i\rangle$,
is directly obtained from~(\ref{flucp}).
To~make the calculation of the expectation value of the
potential analytic, we choose the~internucleon potential of
such a~form as in~\cite{fel95},
\bel{poten}
V=\sum_k \, u_k \, \sum_{i<j} \Big(w_k+ (1-w_k) P^M_{ij}\Big)
\exp\Big(-\lambda_k^{-2} \, (\x_i-\x_j)^2\Big) \, ,
\ee
where $P^M_{ij}$ is the Majorana operator exchanging spatial
coordinates of the particles~$i$ and~$j$.
The~sum over $k$ extends over the repulsive and attractive
contributions to the potential.
The values for $u_k$, $\lambda_k$ and $w_k$ are listed in
\cite{fel95}.
With~(\ref{poten}), we~exclude the Coulomb potential as
in~\cite{fel95}, but then we only consider
the light to medium nuclei.
The expectation value of the potential~(\ref{poten}) with
respect to the wave function~(\ref{trialwf}) is then
\bel{epotexp}
 \langle V \rangle =\sum_k u_k \sum_{i<j} {\cal N}^2 \Big[
 \exp{(-\lambda_k^{-2}(\x_i-\x_j)^2)} \left(w_k + (1-w_k)
P^M_{ij}\right) \Big]
\ee
where the meaning of the square brackets is such as below
Eq.~(\ref{skewalt}).  On~defining the matrices~$\Lambda^{ij}$
and $A_P$ by the relations, respectively,
$\Lambda^{ij}_{nm}\r_n\cdot \r_m=(\r_i-\r_j)^2$ and
$A^P_{nm}= A_{P(n)\,P(m)}$, where $P$ exchanges indices $i$
and~$j$, we obtain
\bel{partpot}
\Big[\exp(-\lambda_k^{-2} \, (\x_i-\x_j)^2) O
\Big]={{\pi^{3N/2}} \over
{ \left( {\rm det}
(A+A^{O \ast}+\lambda^{-2}_k \, \Lambda^{ij} ) \right)^{3/2}}}
\, \exp(\varphi^{ij,k}_O) \, .
\ee
The argument of the exponential on the r.h.s.~of
(\ref{partpot}) for the two cases of the operator~$O$ is
\bel{vexpoe}
\varphi^{ij,k}_{O=1}=-{(\r_i-\r_j)^2\over 2 \lambda_k^{2}} \,
\Big(\Lambda^{ij} \, (2 \, {\rm
Re}A+\lambda^{-2}_k \, \Lambda^{ij})^{-1} \, {\rm Re}A
\Big)_{ii} \, ,
\ee
and
\bel{vexpopi}
\varphi^{ij,k}_{O=P}=
-{ (A+A^{P\ast}+\lambda^{-2}_k \, \Lambda^{ij})^{-1}_{nm}\over
4} \, \q^P_n \, \q^P_m -
{(A+A^{P\ast})_{nm}\over 4} \, \s^P_n \, \s^P_m \, ,
\ee
with $\s^P_n=\r_n-\r_{P(n)}$ and
$\q^P_n=\p_n-\p_{P(n)}-i(A-A^{P\ast})_{nm} \s^P_m$.

At first glance, the calculation of $\langle V\rangle$ for
non-diagonal width-matrix~$A$ scales with the number $N$ of
particles as~$N^5$;
for each of $N(N-1)/2$ particle pairs, one has to invert
a~$N\times N$ symmetric matrix.
Fortunately, for each of the particle pairs the~matrix that
needs to be inverted differs from the matrix
$2\, {\rm Re}\, A$ only in $4N-6$ elements.  Thus, for each of
these pairs one
can use information from inverting the matrix ${\rm Re} \, A$,
and the calculation of~$\langle V \rangle$ scales then only
with the particle number as~$N^4$.

Generally, one would want the trial wave function to be
antisymmetrized.  However, the~width matrix~$A$ nondiagonal in
particle indices introduces such a~number of parameters, that
the antisymmetrization ceases to be feasible for the particle
number larger than $\sim 16$.  Thus, we~resort to the Pauli
potential to simulate the effects of antisymmetrization.
The Pauli potential acts only between particles with the same
spin and isospin, and it is chosen proportional
to the Majorana operator, $V_{Pauli} = u_3 \, P_{ij}^M$.
The~latter is motivated by the fact that the expectation value
of $P^M$ for the product wave function is proportional to the
square of the scalar product of single-particle wave functions,
$\langle P_{ij}^M \rangle = | \langle \psi_i | \psi_j \rangle
|^2$, where $\psi_{i,j}$ are the single-particle wave functions of
the two particles.  When one wave function approaches another,
the system reacts with a repulsive force.  The Pauli potential
is added as the third component to~(\ref{poten}) ($w_3 = 0$,
$\lambda_3^{-1} = 0$), and the value of $u_3$ is adjusted to
best fit the properties of ground-state nuclei.  Unfortunately,
such $u_3$ depends quite significantly on the mass.  Using
parameter
set $B1$ from \cite{fel95} (originally from \cite{BriBo})
in~(\ref{poten}), we~obtain e.g.~optimal $u_3 = 70$~MeV for
$A=12$ and $u_3 = 200$~MeV for $A=80$.  (This would have been
likely partly alleviated if we included the Coulomb potential.)

As an example, in the obtained ground state of~$^{12}$C three
four-fold degenerate centroids position themselves at the
corners of an~equilateral triangle in configuration space.
The~average associated particle momenta are zero.  For~the
diagonal width matrix and the~$B1$ interaction, the~minimum
internal energy (defined as the difference between the total
energy and the energy of the center of mass motion) of
$E_{int}= -87.1$~MeV is obtained for Re~$A_{ii}=0.33$~fm$^{-2}$
which gives~$R_{rms}=2.32$~fm.  The~obtained $^{12}$C nucleus
is stable against break up into three $\alpha$ particles which
have a~ground state energy of~$E_{int}= -28.1$~MeV.  (Note that
the~$^{12}$C energy does not contain the Coulomb energy
estimated
at~$E_C = 11.4$~MeV using the formula~(61) in~\cite{fel95};
the~value of~$R_{rms}$ includes the spatial extent of
a~proton~\cite{fel95}.)  For~the width matrix with off-diagonal
elements, a~lower minimum of~$E_{int}=-95.4$~MeV is obtained
for Re~$A_{ii}=0.30$~fm$^{-2}$, and
Re~$A_{ij}=-0.038$~fm$^{-2}$, if~$j$ is within the same cluster
as~$i$ or has the same spin and isospin directions as~$i$, and
Re~$A_{ij}=0$, otherwise.  There is a~freedom, in~the latter
case, of adding
a~constant to all elements of~$A$ that only changes constraints
on the c.m.~motion.

For even-even nuclei of mass larger than carbon, our Pauli
potential favors differences in average single-particle momenta
over differences in centroids in the ground state.
The~centroids for these nuclei become identical while the
momenta get distributed in momentum space in the groups of
four.
For~example, in~the ground state of $^{16}$O the momenta are
placed at the corners of a tetrahedron in momentum space.

\section{SOLUTION OF THE EQUATIONS OF MOTION}
\label{solution}

We first discuss differences in the dynamics of isolated light
fragments, for correlated and uncorrelated trial
functions.  We~then investigate fragment production
within
the present dynamic description, when a~compressed and excited
nuclear system expands and when nuclei collide at low energies.
Some of our results are quite unexpected.

Since the differences in the dynamics for
correlated and uncorrelated trial functions are expected to be
the largest for the lightest of fragments, we~investigate the
dynamics of
an~isolated deuteron and of an~isolated
$\alpha$~particle, illustrated in Figs.~1 and~2, respectively.
As~the deuteron is unbound for all the interactions listed
in~\cite{fel95}, we~use the Volkov~1~(V1)~\cite{vol65} with
$u_a = -104.5$~MeV in the deuteron case.
Fig.~1a displays the evolution of total $E_{tot}$, kinetic
$E_{kin}$, potential $E_{pot}$, and total internal $E_{int}$
energies, for a~deuteron initialized in the~state of the lowest
internal energy.  Either the dynamics for
a~correlated (labeled~$c$) or uncorrelated (labeled~$u$)
trial-function is followed.  In~the
case of the correlated dynamics, both the total and the
internal
energies remain constant as a~function of time.  However,
in~the
case of the uncorrelated dynamics, the~internal energy
increases, at a~cost of the center-of-mass energy.  At~$t
\sim 12$~fm/c the deuteron becomes, in~effect, unbound and
it remains so thereafter.  The~different evolution is
associated
with the behavior of matrix elements, as~illustrated in
Fig.~1b.
In~the absence of correlations between particles, i.e.\ $A_{12}=0$
in
Eq.~(\ref{deuwf}), the~delocalization of the deuteron center
of mass couples to the~delocalization of the internal state.
In~particular, a~reduction in Re~$A_{cm}$ requires a~reduction
in Re~$A_1$
and this implies a~reduction in~Re~$A_{rel}$.
By~contrast, in~the correlated
dynamics, the~width for the center of mass,
related to Re~$A_{cm}
= \sum_{ij}$~Re~$A_{ij}$, behaves like the width of the
Gaussian
wave-packet for a~particle with twice the nucleon mass.  While
the element~$A_{cm}$ decreases with time, it does so because
the off-diagonal
element $A_{12}$ of the matrix increases in magnitude.
The~element $A_{rel}$, cf.~below~(\ref{deuwf}), stays constant.

As~we have already demonstrated, for the uncorrelated dynamics
an~unphysical
exchange of energy occurs between the internal degrees of
freedom and the center of mass motion.  When the magnitude of
the internal energy exceeds the energy of localization of the
center of mass, the~coupling may cause unphysical oscillations
for the ground state.  This is shown for the~$\alpha$ particle
in~Fig.~2.  In~the case of a~correlated wave function,
the~width
for the center of mass behaves as the width of a~Gaussian
packet
for a~particle with four times the nucleon mass.  The~internal
part of the wave function does not change with time.  In~the
case of an~uncorrelated wave function initialized in the
lowest state of
internal energy, the~matrix element and, correspondingly,
the~width for the center of mass oscillate.  An~exchange of
energy, back and forth, occurs between the internal and
center-of-mass degrees of freedom.  If~the uncorrelated wave
function is initialized in the lowest state of total energy,
the~internal and center-of-mass energies stay constant.
The~internal wave function does not
vary with time,
but nor does vary the center-of-mass wave-function,
with the center of mass never getting delocalized.  Needless to
say that in that case the internal energy is higher than in the
ground
state.  The~examples in Figs.~1 and~2 show
the benefits
of using the correlated over the uncorrelated wave functions in
the fragment description.

Turning now to fragment production, we start out by exploring
the
situation where a highly excited system formed in the central
heavy-ion collisions decays into vacuum.
For such a~system we
expect a~reduced importance of the antisymmetrization that is
missing from our equations.
We assume that internal degrees of freedom of the system
are, generally,
equilibrated, allowing only for a~variable strength of
the radial flow.  We investigate the dynamics with a~variable
width matrix, either restricted or not to the~diagonal form
and, further, the~dynamics with a~static width, i.e.~classical.
More significant and surprising differences are found between
the quantal
and classical dynamics than between the dynamics with different
variable width matrices.

To simulate the excited system ($A=20-80$), we
distribute centroids randomly within
a~spatial volume of radius~$R$ and in the momentum according
to a~finite-temperature Fermi distribution ($T=5-12$~MeV).
To~account for the flow, we add to the momentum of each
particle a~component proportional to the position vector
relative to the overall center of mass.
The~proportionality constant determines the amount of flow
energy in the initial state ($E_{col}=0-25$~MeV/nucleon).
The~width matrix is initialized as a real multiple of the
unit matrix.
Spatial representation of one of such initial states is given
in~Fig.~3a.
In~the particular case $A=80$ and $R \simeq 5$ fm.  The~rms
radii of individual packets are equal to~1.9~fm.

In~the case of
a~dynamic matrix, whether or not restricted to a~diagonal
form, the~excited system, initialized as above, emits
a~number of single nucleons in~the course of time.
The~number of emitted nucleons generally
increases with the energy of the system.  However, at~no
particular energy, for the studied $A=20-80$ systems, {\em any
emission of IMF or even of $\alpha$ particles is observed}.
This appears to be true irrespective of how we
divide the excitation energy into collective and thermal.
Examples of the late-stage distributions of centroids in space
are given in Figs.~3(c) and~3(d) for the initial
net excitation energy of 14 and 26~MeV/nucleon, respectively.
For the particular
initial states, the centroid distributions at the respective
time are not distinguishable by eye between the evolutions with
and
without correlations.  We~found that frequently to be the
case
for the systems initialized in the manner discussed above.
While some centroids appear
close to each other in Figs.~3(c) and~3(d), at a~distance
from the main residue, the~respective wave-packet widths are so
large that the packets will separate from each
other, eventually.  At~$t = 300$~fm/c, for the displayed
systems,
all relative energies of the emitted particles are positive.

To check whether the lack of fragment production in the
correlated dynamics might be associated with the initialization
of the dynamics in an~uncorrelated state, we~have carried out
tests using different initializations.  Thus, we have
added to the hamiltonian an~oscillator term $V = v_{osc}
\, \sum_i x_i^2$, to~keep an~excited system from expanding
into vacuum, while allowing the matrix elements of~$A$ to
thermalize, eventually removing this additional potential.
Figure~\ref{vosc} shows the late stage of an~$A=80$ system
initialized using a~narrow oscillator potential.  During the
time of 100 fm/c within the oscillator potential,
the rms values of off-diagonal elements of the matrix~$A$
had saturated, $\langle ( {\rm Re} \, A_{i\ne j})^2
\rangle^{1/2}
\sim 0.7/(A-1) $; for the chosen $v_{osc}$ the system hardly
expanded in the potential, compared to $t=0$.  On~removing the
constraining oscillator potential, we added
collective components to particle momenta,
proportional to the distance from the center.  The~system,
subsequently, emitted a number of {\em single} nucleons and
a~deexcited residue formed at the center,
consisting of~37 nucleons at the time shown in Fig.~\ref{vosc}.
In~tests we changed the extension of the constraining
oscillator potential for the excited nucleus, the~initial
temperature, and the~magnitude of collective energy.  Further,
we initialized excited nuclei without any initial constraining
potential, just assigning random gaussian values to the
off-diagonal and diagonal elements of~$A$, using the
thermalized values
from the oscillator potential as a~guidance.  Consistently, in
all tests,
the~released nuclei emitted, in the course of their evolution,
a~number of single nuclei but never any IMF or even an~$\alpha$
particle.  Thus, the~particular feature does not depend on the
off-diagonal terms of~$A$ being zero or finite in the initial
state of an~excited nucleus.

Clearly, in the past, the production of IMFs and light clusters
has been observed in the QMD calculations.  The~QMD limit
corresponds to taking $\hbar \rightarrow 0$, or to suppressing
the width dynamics in our equations.  Indeed, when taking
$\hbar$ reduced by a~factor of (5--10), or the width dynamics
slowed down by a such a~factor, we begin to observe the cluster
production.  Figure~3(b) shows centroids for a~system
evolved from the initial state shown in
Fig.~3(a), using a~static width, for the same available energy
(given the frozen
width) as the system in~Fig.~3(c).  At a~time $t=300$~fm/c
in~Fig.~3(b), two IMFs as well as two dinucleons are seen and
they will remain stable.
Additional small clusters have been emitted before $t=300$~fm/c
and
left the displayed spatial region.  As~the reason for the lack
of cluster production in the calculations with a~dynamic width
matrix (whether diagonal or not) emerges, in our tests,
the~spreading of wave functions towards
the emission time.  After packets get delocalized,
the~interaction
is not capable to contract them back into fragments.

Our results on the cluster production for the dynamic width
matrix may seem in contradiction to the FMD
results~\cite{fel95}
with even multifragmentation events reported in nuclear
collisions.
However, a~scrupulous examination indicates that the clusters,
seen
in the final states of the FMD calculations in~\cite{fel95},
were not formed
during reactions, but were present in the initial states and
survived reactions.
When investigating that particular issue, we simulated
symmetric
collisions of nuclei with different initial structures.  Thus
e.g.\ for the
potential~(\ref{poten}) and our Pauli potential,
the~centroids within the ground state of $^{12}C$ form three
$\alpha$-type clusters of four nucleons each.
On~the other hand, within $^{40}Ca$ the nucleon centroids
situate themselves at the overall center of mass
position;  the~widths for different nucleons take on different
values. In the true ground state for our potential, the momenta
form groups of four in $^{40}Ca$, with the four
particles
being two protons and two neutrons with different spin
directions.  To~illustrate, though, our point on reactions, we
shall
displace slightly the momenta from the identical values in
$^{40}Ca$ in the initial state of a~reaction, making sure that
change in the overall binding
energy is negligible.  Notably, inclusion of any kind of
spin-isospin dependent interaction would break
the momentum sub-clusters in nuclei of $A\le 40$, anyway.
Then, when considering the $^{12}C+^{12}C$ reaction, we shall
deal
with six sub-clusters in initial state, and with none in the
$^{40}Ca+^{40}Ca$ case.

Figure~\ref{C+C} shows the initial and late states of an
exemplary $^{12}C+^{12}C$ reaction at the beam energy
of 29~MeV/nucleon.
While one nucleus got highly excited in the reaction, the
other has fragmented into three $\alpha$ particles.
Each of the $\alpha$ particles is excited at the displayed
time, however remaining below threshold for particle
emission.

Figure~\ref{Ca+Ca} shows three stages of a~35~MeV/nucleon
$^{40}Ca+^{40}Ca$ reaction with no sub-clusters in the
initial state.
At~the first of the times shown, the~wave functions of the
two nuclei just started to overlap.
At~the second of the times, a~transient residue that formed is
maximally
spread out.  At~the third of the times, the~outcome of the
collision is, essentially, determined and
only a~residue and some single
nucleons are seen.
No~clusters are emitted from this collision.

The examples presented in Figs.~\ref{C+C} and~\ref{Ca+Ca} are
typical ones: for none of the
studied initial states, we have observed, for the dynamic
width, the emission of IMF's or of
$\alpha$ particles that were not present in the substructure of
an~initial
state.  This has been the case whether we included or excluded
the correlations.  Again,
as a~reason for the absence of cluster formation for
the dynamic width, we find that, at times when the system
reaches a~density when clusters are expected to form, the width
of the wave packets has grown so large that the interaction
between different wave packets is too weak to force the width
of these packets to shrink, and their centroids to approach
each other enough to form a~nucleus (see
also~\cite{cho96}).  When we suppress the dynamics of
the matrix~$A$, new clusters form and emerge from
the~reaction region.

The values of the widths are not necessarily unphysically
large, as nucleons are expected to get delocalized with the
reaction progress.
In~reality, though, when broad wave packets of different
nucleons overlap in space, the~interaction between the nucleons
should be generally able to clump two or more nucleons, within
a~distance comparable to the interaction range, into a~cluster.
The~wave function in terms of broadened Gaussians (whether or
not diagonal in the nucleon coordinates) does not allow for
that.

\section{Summary and Conclusions}
\label{summary}

We have investigated the time evolution of nuclear systems,
in terms of correlated and uncorrelated Gaussian wave
functions,
following from the time variational principle.
As~an~interaction in the Hamiltonian, we utilized a~Volkov-type
potential.  Upon determining that an~antisymmetrization
of the correlated wave function would not be
feasible beyond relatively small systems, we employed
a~Pauli type potential to simulate the effects of
Pauli principle.

For uncorrelated trial wave functions, the internal state
cannot be localized without localizing the center of mass.
Thermal estimates indicate that this could suppress fragment
emission.  In~dynamics, an~unphysical energy exchange occurs
between the center of mass and the intrinsic motions.  These
deficiencies are absent when using the {\em correlated} wave
functions.  We have explicitly demonstrated an
improvement, in terms of correlated wave functions, in the
description of deuterons and alphas.

Contrary to expectations, the~inclusion of correlations has not
improved the description of cluster production in the explicit
simulations of heavy-ion reactions or in
situations characteristic for the reactions.  Either in
correlated or uncorrelated Gaussian dynamics, clusters are
{\em only} produced when they are present in the substructure
of an~initial state.  This appears to be true, in the
correlated dynamics, whether or not the initial state
is correlated.  The~absence of new clusters is associated with
a~large spreading of the wave function at the~time when
a~reacting system expands and the new clusters are expected to
form.
The~spreading in relative coordinates is present even in
the correlated wave function, despite of the fact that,
for a~correlated wave function, the~relative spreading may
evolve independently from the spreading
for the center of mass for any of the potential fragments.
The~interaction is too weak to pull back the wave function to
a~size appropriate for a~fragment.
In~classical dynamics fragments are produced.

While there is nothing unphysical
in the spreading of the wave function as such in the
simulations,
in~reality
the~interaction would be capable of creating correlations in
the wave function over distances of the interaction
range.  E.g., when considering the two-nucleon correlation
function, with time the function should develop into
a~spreading
long-range component, weighted with a~certain probability, and
a~more stable short-range component corresponding to
the forming fragments.  The~specific parametrization of the
wave
function, once the wave function has spread, does not allow
for the development of the short-range component.

On a~general level our results show that the dynamics is
important
in fragment production.  The~spread of the wave packets at
emission indicates that an~emitting system can gain
the packet delocalization energy.  On~the side,
within~AMD
the width dynamics is suppressed and an {\em ad hoc} term is
added to the Hamiltonian to account for the delocalization at
emission.
Within~QMD, the location energy
is disregarded at every stage, including
the initial ground-state nuclei and the final state.
Practically, the~procedure in AMD amounts to
a~change in the effective interaction, compared to QMD.
Within either the~approaches, the~fixed width of a~wave packet
extends the range of the two-particle interaction and may, in
fact, act to suppress fragment formation.

Recent results from FMD with short-range
correlations~\cite{fel96} show some fragments in the final
state of reactions, in~contrast to the present calculations
(or~\cite{cho96}).  This can be attributed to a~lesser
spreading of
the wave packets towards emission time, due to the~combination
of the effects of short-range correlations and the assumed
spherical shape of packets.  Given that this lesser spreading
is unrelated to the physics of fragment production, we
believe that it is not a~solution to the
the dilemma at hand.
In~particular, we expect too low fragment yields, from
the calculations with short-range correlations,
compared to experiment.

We conclude that, in a~successful description of fragment
production, the wave function or, more generally,
the~model density matrix must a have a flexibility
to change over distances comparable to the interaction range,
at a~time when the fragments are formed. One possibility within
the
dynamics is to keep the nucleon wave-packet width static and
comparable to the interaction range, and to account for
a~quantal spreading with time through a stochastic decision
process~\cite{ohn95}.
This can be nominally derived through the reduction of the wave
function space to the space spanned by the wave packets of
constant width, which results in a~residual force
associated with the kinetic-energy part of the hamiltonian.
The~present authors have been exploring a~replacing of
the~single wave
packet for every nucleon by a~superposition of packets.
Unfortunately,
this looses the inherent simplicity of~FMD.  The~high hopes
associated with the FMD approach
cannot be directly realized as far as fragmentation is
concerned.

\acknowledgements
The authors benefited from discussions with H.~Feldmeier,
J.~Schnack, and M.~P\l oszajczak.  Suggestion of H.~Feldmeier
to investigate an~initial state with long-range correlations is
appreciated.
This work was partially supported by  the Deutsche
Forschungsgemeinschaft and by the National Science Foundation
under Grant No.~PHY-9403666.

%\pagebreak

\newpage

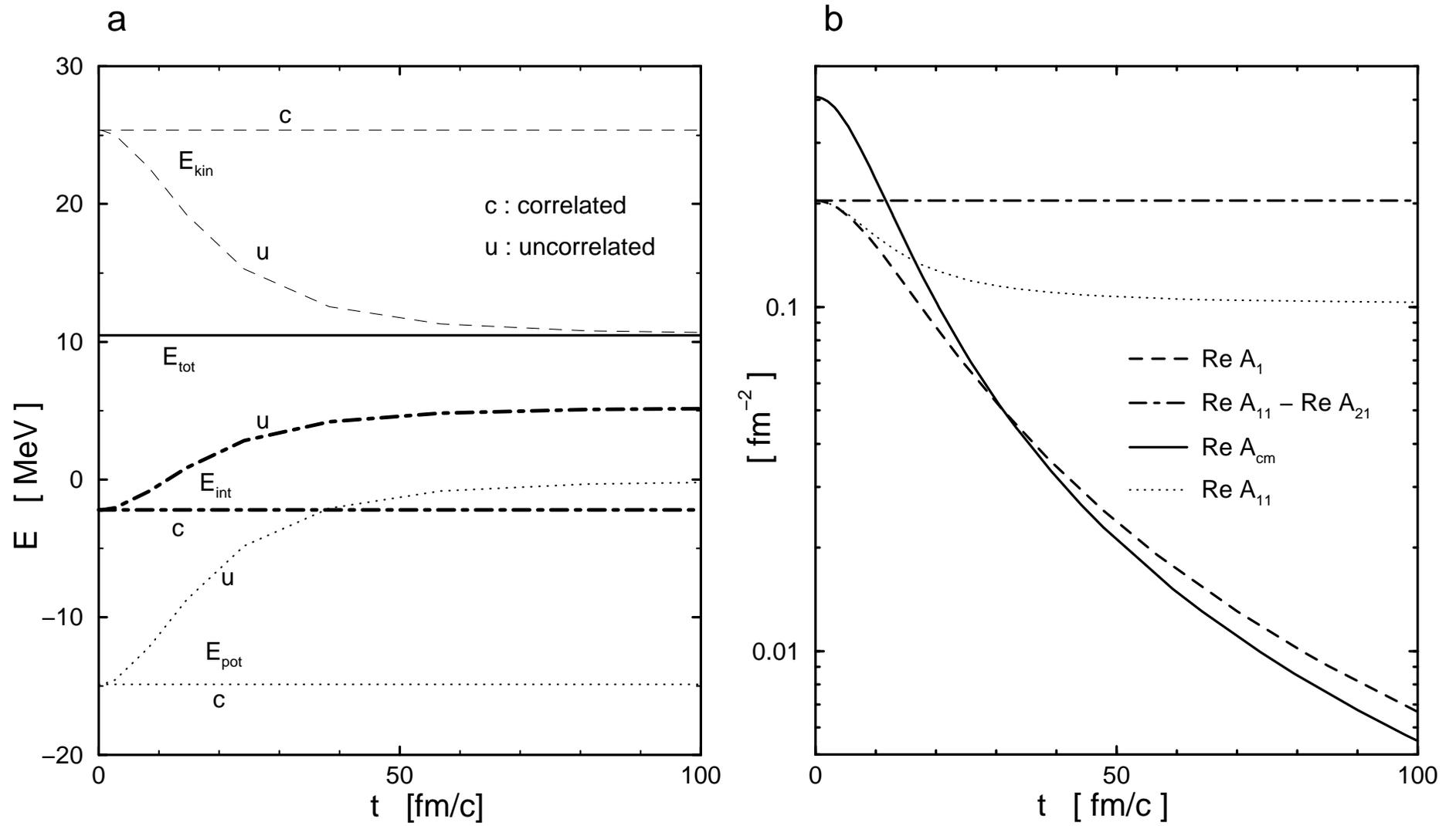
\begin{figure}
\caption{
Evolution for the internal ground state of a free deuteron.
(a)~The~dashed, solid, dash-dotted, and dotted lines show,
respectively, the~evolution of the kinetic, total, internal,
and potential energies
in correlated~($c$) and and uncorrelated~($u$) dynamics.
(b)~Evolution of the elements of the width matrix.  The~dashed
lines shows the evolution of the element in the case of
an~uncorrelated
wave function.  The~dash-dotted, solid, and dotted lines show,
respectively, the~evolution of the internal, center-of-mass,
and diagonal elements in the case of a correlated wave
function, cf.~Eq.~(\protect\ref{deuwf}) and text below.
}
\end{figure}

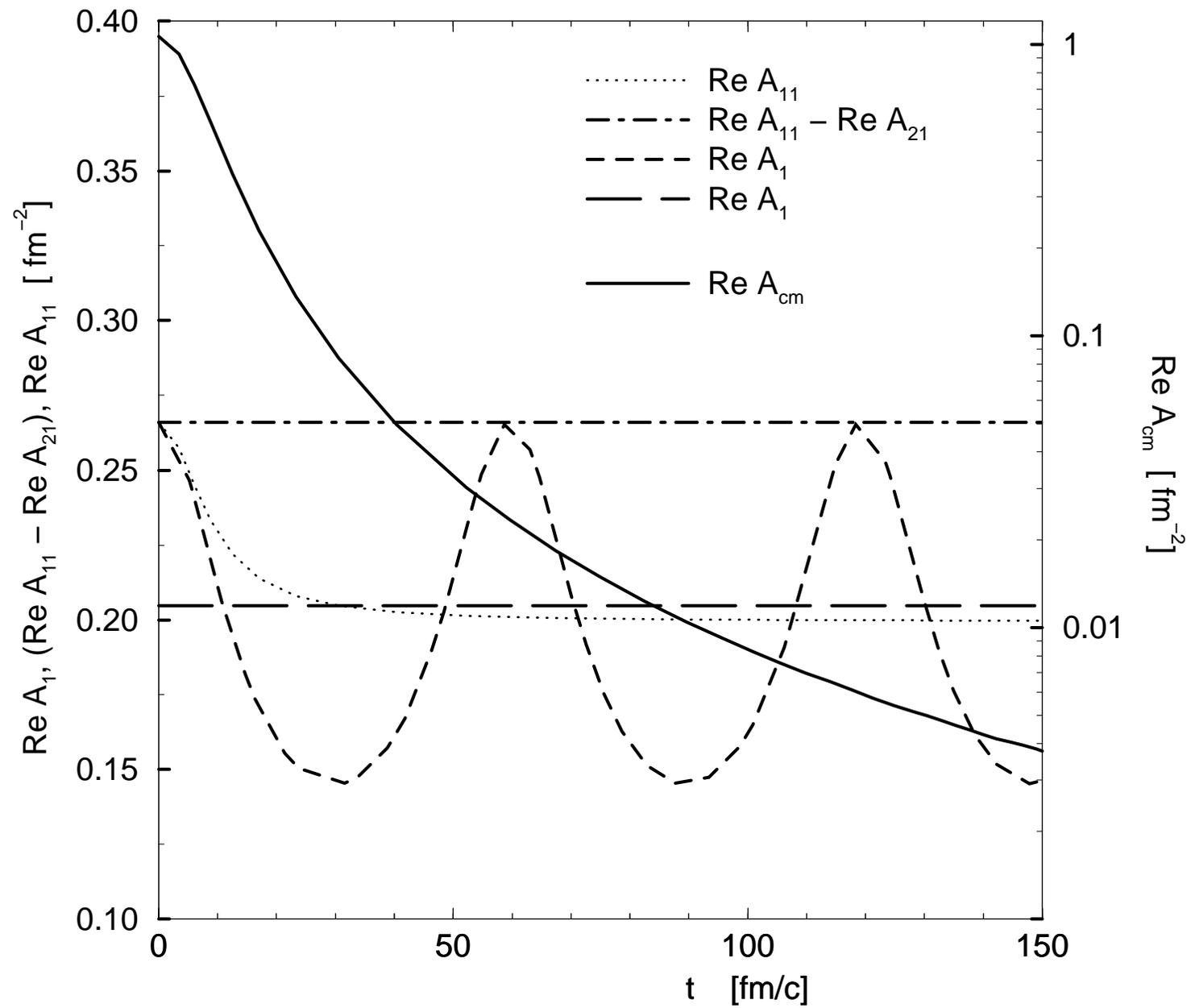
\begin{figure}
\caption{
Evolution of the elements of the width matrix~$A$ for
an~$\alpha$
particle initialized in its ground state.  The~solid,
dash-dotted, and dotted lines show, respectively, the~evolution
of the center-of-mass, relative between-two-nucleons, and
diagonal elements, in~the case of a~correlated wave function,
cf.~Eqs.~(\protect\ref{deuwf}) and~(\protect\ref{expon}).
The~long-dashed line shows the evolution of the element in
the case of
an~uncorrelated wave function initialized in the lowest state
of internal energy.
The~short-dashed line shows the evolution of the element
in the case of
an~uncorrelated function initialized in the lowest state of
total energy.
}
\end{figure}

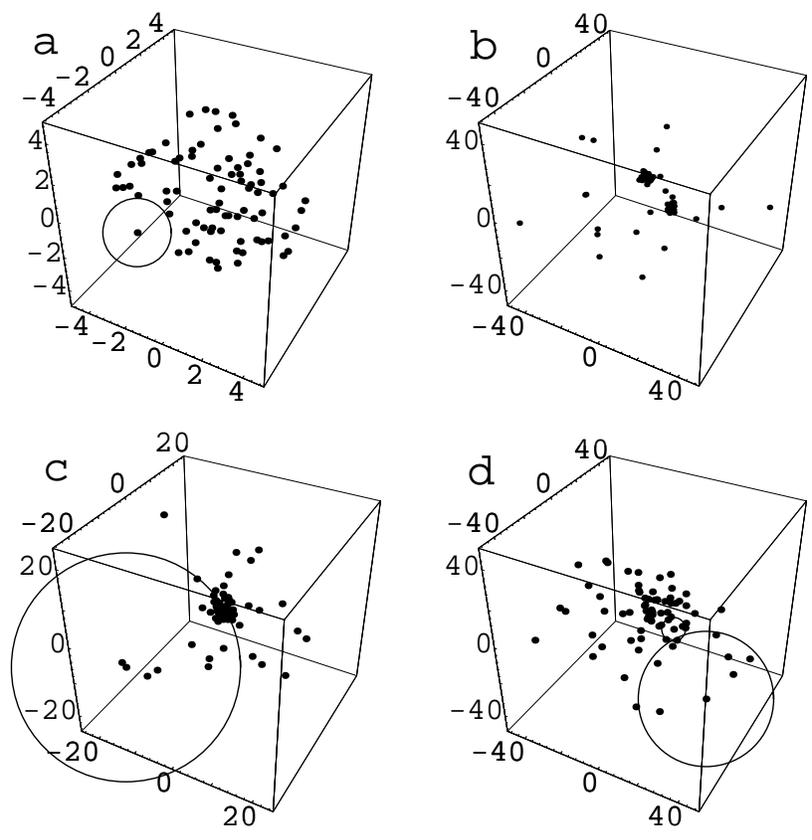
\begin{figure}
\caption{
Particle centroids (dots) in the configuration space for
$A=80$, in the initial state~(a) and at $t=200$ {\rm fm/c}
for the static wave-packet width~(b), and for the dynamic
width matrix~(c) and~(d). In~(c), the initial flow energy
is lowered, compared to (b) and~(d), by the initial energy
content in the localization of the wave packets.  The
circles indicate rms radii of the most and least localized
Gaussians.  The~radii of the packets in the case (b), and
the radii of the most localized packets in the case (c),
are comparable to the radii of the dots representing centroids.
The~axes show distances in fm.
}
\end{figure}

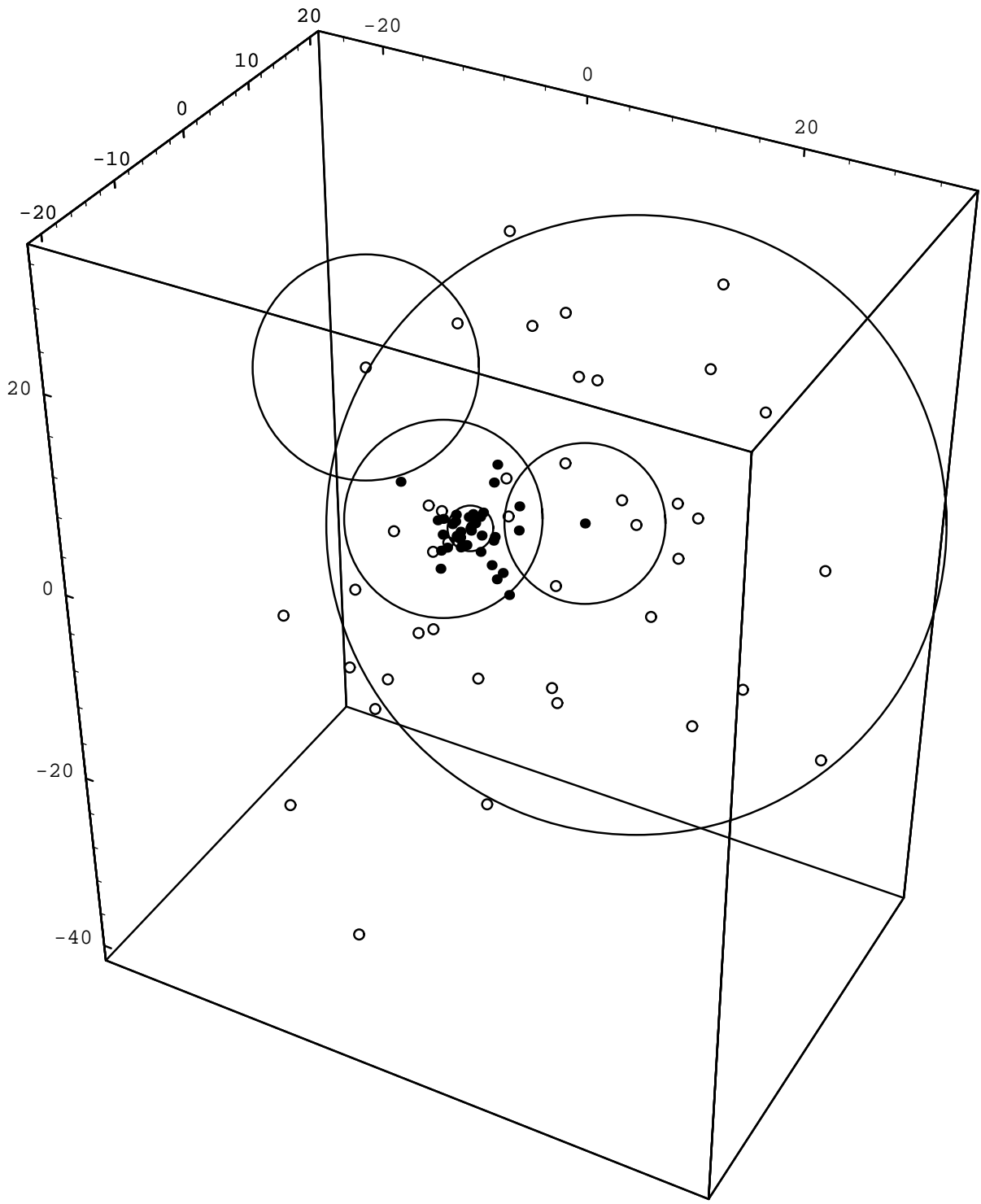
\begin{figure}
\caption{
Late stage ($t=300$~fm/c) of an $A=80$ system initialized at
a temperature $T=8$~MeV and $E_{col}=0$ in a narrow
oscillator potential $V = v_{osc}
\, \sum_i x_i^2$, where $v_{osc}=1$~MeV/fm$^2$.  At a time
$t=100$~fm/c, the oscillator potential was removed, and
particles given collective outward velocities corresponding to
net
$E_{col}= 7$~MeV/nucleon.  The centroids for emitted and bound
particles (negative and positive removal energies,
respectively) are indicated with small open and small filled
circles, respectively.  The large circles indicate widths, from
the diagonal elements of~$A$, for most localized and
delocalized bound and emitted particles.
}
\label{vosc}
\end{figure}

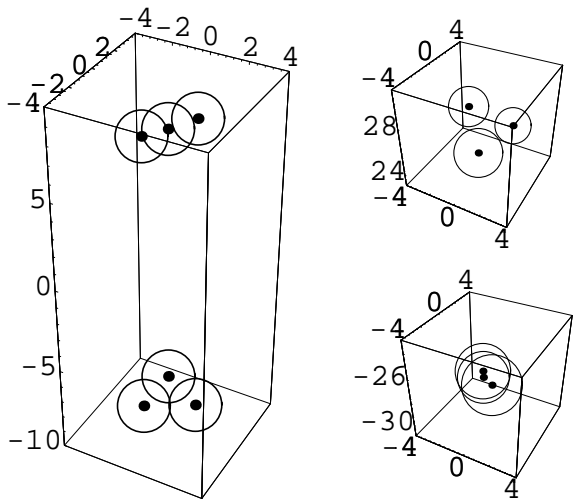
\begin{figure}
\caption{
Initial (left side) and late ($t =300$~fm; right side) states
of a~$^{12}C+^{12}C$ reaction at 29~MeV/nu\-cle\-on. The~beam
axis is directed vertically, dots represent the centroids of
nucleon packets, and circles show the packet rms radii.  Edges
of
the spatial boxes show distances along the carthesian axes
in fm.
}
\label{C+C}
\end{figure}

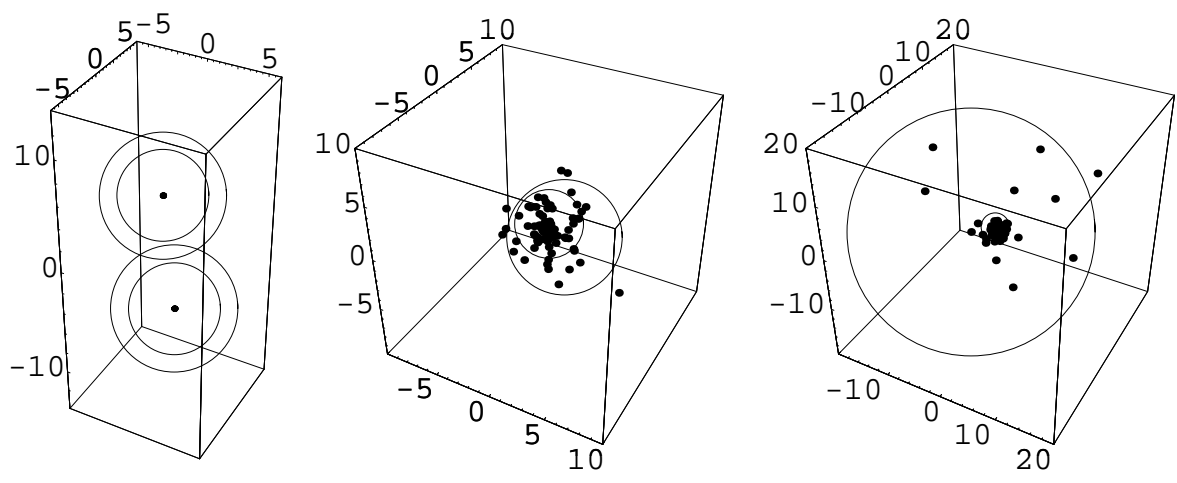
\begin{figure}
\caption{
Three stages of a~35~MeV/nucleon $^{40}Ca+^{40}Ca$ reaction
in the configuration space.  From left to right,
the~boxes show the reaction, respectively, at
$t=20$ fm/c, $t=100$ fm/c, and $t=300$ fm/c.
The~dots indicate centroids, while the circles indicate the rms
radii of the least and most localized Gaussians.  Box edges
show the distances along the carthesian axes in~fm.
}
\label{Ca+Ca}
\end{figure}

\end{document}